\documentclass[12pt,a4paper,notitlepage,aps]{revtex4}
\usepackage{graphicx}
\usepackage{latexsym,natbib,pxfonts}
\newcommand\beq{\begin{equation}}
\newcommand\eeq{\end{equation}}
\newcommand\bea{\begin{eqnarray}}
\newcommand\eea{\end{eqnarray}}
\begin{document}
\title{SU(N) Irreducible Schwinger Bosons}
\author{Manu Mathur\footnote{manu@bose.res.in}, Indrakshi Raychowdhury\footnote{indrakshi@bose.res.in}}
\affiliation{ S. N. Bose National Centre for Basic Sciences\\ Block JD, Sector III, Salt Lake, Kolkata- 700098, India}
\author{Ramesh Anishetty\footnote{ramesha@imsc.res.in}}
\affiliation{ The Institute of Mathematical Sciences\\
CIT-Campus, Taramani, Chennai, India}
\begin{abstract}
We construct SU(N) irreducible Schwinger bosons satisfying certain U(N-1) constraints which 
implement the symmetries of SU(N) Young tableaues. As a result all SU(N) irreducible representations are 
simple monomials of $(N-1)$ types of SU(N) irreducible Schwinger bosons. 
Further, we show that these representations are free of multiplicity 
problems. Thus all SU(N) representations are made as simple as SU(2). 
\end{abstract}
\maketitle
\section{Introduction}
\label{intro} 
The Schwinger construction of SU(2) Lie group and all its representations 
in terms of two simple harmonic oscillators or equivalently Schwinger bosons
\cite{schwinger} is well known. Due to its simplicity, this construction has been 
widely used in various branches of physics like nuclear physics \cite{np}, 
strongly correlated systems \cite{scs}, 
supersymmetry and supergravity algebras \cite{susy}, lattice gauge theories \cite{lgt}, 
loop quantum gravity \cite{lqg} etc..
The three novel features of SU(2) Schwinger boson construction are 
its completeness, economy and simplicity. More precisely, the Hilbert space created 
by the two Schwinger oscillators is isomorphic to the representation space of SU(2) group 
(see section \ref{su2}). 
Thus the Schwinger boson representation of SU(2) group is complete (all SU(2) representations 
occur) as well as economical (every representation occurs once). Further, this construction 
is also simple as all SU(2) representations are given in terms of monomials (not polynomials) 
of Schwinger bosons. 
It is well known that all the above desired features of economy, completeness and elegance associated 
with SU(2) group are lost when we consider mixed representations of SU(3) or higher SU(N) groups. There has been 
considerable work in the past in these directions \cite{mosh,gel,apw,sharat,sharat2,bied,mm1,mu1,rmi}. 
In fact, following the work of Gelfand \cite{gel}, an explicit realization of a group G leading to 
its representations which are complete and without multiplicities is known as a model or Gelfand model 
of G in mathematics literature and is a subject of considerable interest \cite{bied}. 
The reason for multiplicities in SU(N) representations is that SU(N) group requires at least $(N-1)$ fundamental 
representations to get its all other irreducible representations. This immediately implies 
existence of certain non-trivial SU(N) invariant operators (see the references 
section \ref{su3},~ \ref{su4} and \ref{sun} and references therein). 
Any two states which differ by an overall presence of such invariants will transform in the 
same way under SU(N). This leads to the problem of multiplicity which in turn 
makes the representation theory of SU(N) ($N\ge 3$) much more involved than 
SU(2). The standard way to project out these invariants is by using SU(N) Young tableaues. 
The symmetrization and anti symmetrization of SU(N) indices along the rows and columns of SU(N) Young 
tableaues remove these invariants leading to all SU(N) irreducible representations. 
However, the symmetrization and anti-symmetrization operations, in turn, 
make the representations of even the simplest SU(3) group extremely complicated 
(see section \ref{su3}). As N increases the representations become more and more 
complicated. This renders them useless for any practical application. Note 
that this is unlike SU(2) case where Schwinger bosons creation operators commute amongst themselves 
and hence have built in permutation symmetry of SU(2) Young tableaues along its row (see section \ref{su2}). 
In this work, we define 
SU(N) irreducible Schwinger bosons (henceforth we call them SU(N) ISB) with built in symmetries 
of SU(N) Young tableaues. As a result in terms of these SU(N) ISB all SU(N) representations are monomials like in simple SU(2) 
case. Further, we show that all SU(N) invariant operators constructed out of SU(N) ISB trivially annihilate the 
above SU(N) representation states. Hence, like SU(2) case, SU(N) ISB representations are multiplicity free. 
Another feature of the construction is that it iterative in nature. The results obtained for creation operators for 
SU(N) get carried over to SU(N+1) without any change (see section \ref{su4} and \ref{sun}). In fact, the present 
work is SU(N) generalization of SU(3) work \cite{rmi}. 
The plan of the paper is as follows. It contains four sections on SU(2), SU(3), SU(4) and SU(N) 
ISB respectively so that the presentation remains transparent and self 
contained. In section \ref{su2} we start with SU(2) Schwinger boson construction 
briefly to highlight all its special and simple features. In section \ref{su3} we describe the earlier 
work \cite{rmi} on SU(3) irreducible Schwinger bosons and then reformulate the problem in a language which can 
be directly generalized to SU(N). Before proceeding to general case, we note that unlike SU(2) and SU(3) the 
higher SU(N) ($N \ge 4$) groups 
have fundamental representations which are not N-plets leading to different types of group invariants.
In fact, this is the reason why the SU(3) irreducible Schwinger boson construction in \cite{rmi} 
can not be directly generalized to higher SU(N) and requires change of language. 
Therefore in section (\ref{su4}) we discuss the SU(4) representations explicitly. 
As we will see this SU(4) section makes the transition from SU(2), SU(3) to SU(N) easy and 
smooth. In section \ref{sun} we construct all the $(N-1)$ SU(N) ISB and 
show that all SU(N) representations in terms of ISB are monomials without multiplicities. 
We conclude the work with a brief discussion on parallels between the constraints in the 
present work and Gauss law constraints in quantum gauge theories \cite{sharat2}. In fact, 
the constraints leading to the Hilbert spaces containing SU(N) representations without multiplicities are 
analogous to the Gauss law constraints leading to the physical Hilbert space containing states without 
gauge multiplicities. 
\section{SU(2) Schwinger Boson Representations}
\label{su2}
The three generators $\{ J_1,J_2,J_3 \}$ of the group $SU(2)$ satisfy 
the following commutation relation among themselves:
\bea
[J_{\mathrm a},J_{\mathrm b}]=i\epsilon_{{\mathrm abc}}J_{\mathrm c}; \quad {\mathrm a,b,c} =1,2,3. 
\eea
This algebra can be realized in terms of a doublet of Harmonic oscillator creation and 
annihilation operators given by $(a^{\dagger 1},a^{\dagger 2})$ satisfying the algebra,
\bea
[a_{\alpha},a^{\dagger \beta}]=\delta_{\alpha}^{\beta} ~~~~~~~, ~~~~~~[a^{\dagger \alpha},a^{\dagger \beta}]=0~~~~~~,~~~~~~[a_\alpha, a_\beta]=0 ~~~.
\eea
In terms of these operators,
\bea
J^{\mathrm a} = \sum_{\alpha,\beta=1}^{2} a^{\dagger \alpha}\left(\frac{\sigma^{\mathrm a}}{2}\right)^{\beta}_{\alpha}a_{\beta}~~~.
\eea
where $\sigma^{\mathrm a}$ denote the Pauli matrices.
It is easy to check that 
these operators satisfy the $SU(2)$ Lie algebra with the SU(2) 
Casimir: 
\bea 
\label{noc} 
{J} ^2 \equiv {{a}^{\dagger} \cdot {a} \over 2} 
\left({{a}^{\dagger} \cdot {a} \over 2} + 1\right)~~,\eea 
where $a^\dagger\cdot a~(=a^\dagger_1a_1+a^\dagger_2a_2)$ is the total number operator.\\
Thus the representations of $SU(2)$ 
can be characterized by the eigenvalues of the total occupation number operator with the angular 
momentum satisfying,
\bea 
\label{noc1} 
j={\left(n_1+ n_2\right) \over 2} \equiv {n \over 2}~~, \eea
where $n_1$ and $n_2$ are the 
eigenvalues of $a^{\dagger 1} a_1$ and $a^{\dagger 2} a_2$ respectively. \\
An arbitrary SU(2) representation characterized by angular momentum $j= {n \over 2}$ 
is given by the SU(2) Young tableau shown in Figure 1. 
As harmonic oscillator creation operators $a^{\dagger}_{\alpha} (\alpha =1,2)$ commute 
amongst themselves, the simple monomial state: 
\beq
\label{su2irrep}
|\psi\rangle^{\alpha_1\alpha_2...\alpha_{n}} 
=a^{\dagger \alpha_1} a^{\dagger \alpha_2}... a^{\dagger \alpha_{n}}|0\rangle
\equiv O^{\alpha_1\alpha_2 \ldots \alpha_n} |0\rangle. 
\eeq
is symmetric under all possible $n!$ permutations of spin indices $(\alpha_1,\alpha_2, \cdots \alpha_n)$.\\ 
\begin{figure}[h]
\begin{center}
\includegraphics[width=0.5\textwidth,height=0.15\textwidth]
{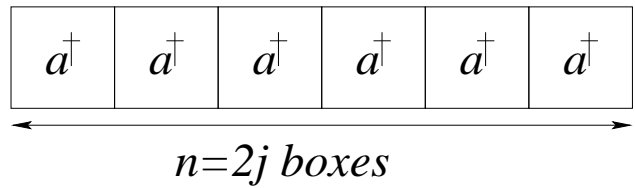}
\caption[su2irrep]{SU(2) Young tableau for the representation $n=2j$. Each SU(2)
Schwinger boson $a^{\dagger \alpha}$ in (\ref{su2irrep}) creates a Young tableau box.} 
\end{center} 
\end{figure} 
Above we have defined $O^{\alpha_1\alpha_2 \ldots \alpha_n} = 
a^{\dagger \alpha_1} a^{\dagger \alpha_2}... a^{\dagger \alpha_{n}}$ for later convenience.
Thus the symmetries of SU(2) Young tableau with n boxes in a row is inbuilt in the construction 
and the states 
$|\psi\rangle^{\alpha_1,\alpha_2,...,\alpha_n}$ in (\ref{su2irrep}) 
form an SU(2) irrep with $j = {n \over 2}$. From now on we 
shall say that SU(2) Schwinger bosons are SU(2) irreducible as they have built in symmetry of 
SU(2) Young tableau. In other words, no explicit symmetrization of spin half indices is required to obtain SU(2) 
irreducible representations. 
As mentioned in section \ref{intro}, the aim of the present work is to define SU(N) irreducible Schwinger bosons 
which have the built in symmetries of SU(N) Young tableaues. Thus all SU(N) representations in terms of SU(N) 
ISB retain the simplicity and elegance of SU(2) representations in (\ref{su2irrep}). 
\section{SU(3) Schwinger Boson Representations }
\label{su3} 
\begin{figure}[t]
\begin{center}
\includegraphics[width=0.5\textwidth,height=0.20\textwidth]
{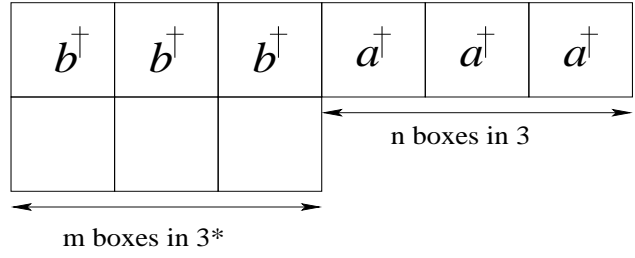}
\end{center}
\caption[NMirrep]{SU(3) Young tableau for the representation $(n,m)$. Each SU(3) irreducible Schwinger bosons $A^\dagger_\alpha(B^{\dagger\alpha})$ creates a single and double Young tableaux box.} 
\label{su3ytp}
\end{figure}

In this section we briefly review the work in \cite{rmi} and then recast it in a new framework which 
is generalizable to SU(N). The details of first part can be found in \cite{rmi}. 
The rank of the SU(3) group is two. Therefore, to cover all SU(3) irreducible 
representations we need two independent harmonic oscillator triplets. 
Let's denote them by $\{a^{\dagger}_\alpha\} \in 3$ and $\{b^{\dagger\alpha}\} \in 3^*$ with 
$\alpha = 1,2,3$. Now the generators of $SU(3)$ group are written as \cite{georgi}:
\bea
Q^{\mathrm a} = a^\dagger {\lambda^{\mathrm a} \over 2} a - b^\dagger {\tilde{\lambda}^{\mathrm a} \over 2} b, 
~~ a = 1,2, \cdots ,8. 
\label{su3sb} 
\eea
In (\ref{su3sb}) $\lambda^{\mathrm a}$ are the Gell Mann matrices for triplet representation, 
$- \tilde{\lambda}^{\mathrm a}$ are the corresponding matrices 
for the $3^*$ representation where $\tilde{\lambda}$ denotes the transpose 
of $\lambda$. 
The defining relation (\ref{su3sb}) implies that under SU(3) $(a^{\dagger})_{\alpha}$ and 
$(b^{\dagger})^{\alpha}$ transform according to $3$ and $3^*$ representations. 
As $Q^{\mathrm a}, ({\mathrm a}=1,2,..,8)$ in (\ref{su3sb}) involve both creation 
and annihilation operators, 
the SU(3) Casimirs are the total occupation number operators of $a$ and $b$ type oscillators: 
\bea
{N}_a = a^\dagger \cdot a, \qquad \qquad {N}_b = b^\dagger \cdot b.
\label{su3cas} 
\eea
We represent their eigenvalues by $n$ and $m$ respectively and the SU(3) vacuum state $(n=0,~m=0)$ by $|0\rangle$.
The representations $(n,m)$ are associated with Young tableau shown in Figure \ref{su3ytp} with n single and 
m double boxes. 
At this stage, we define six dimensional 
Hilbert space ${\cal{H}}_{HO}^{6}$ which is created by six oscillators 
$(a^{\dagger~\alpha}, b^{\dagger}_{\alpha})$ for $\alpha=1,2,3$. The basis vectors can be written as:
$$|{}^{n_1n_2n_3}_{m_1m_2m_3} \rangle \equiv 
\left(a^{\dagger 1}\right)^{n_1} \left(a^{\dagger 2}\right)^{n_2} \left(a^{\dagger 3}\right)^{n_3} 
\left(b^{\dagger}_1\right)^{m_1} \left(b^{\dagger}_2\right)^{m_2} 
\left(b^{\dagger}_3\right)^{m_3} |0\rangle ~~~,$$
or equivalently by: 
\bea
|{}^{\alpha_1,\alpha_2, \cdots \alpha_n}_{\beta_1,\beta_2, \cdots \beta_m}\rangle 
\equiv O^{\alpha_1\alpha_2...\alpha_n}_{\beta_1\beta_2...\beta_m} |0\rangle \equiv
(a^{\dagger})^{\alpha_1} (a^{\dagger})^{\alpha_2} ... (a^{\dagger})^{\alpha_n}
(b^{\dagger})_{\beta_1} (b^{\dagger})_{\beta_2} ... (b^{\dagger})_{\beta_m}|0\rangle ~~ ,
\label{defo}
\eea
with $O^{\alpha_1\alpha_2...\alpha_n}_{\beta_1\beta_2...\beta_m} \equiv (a^{\dagger})^{\alpha_1} 
(a^{\dagger})^{\alpha_2} ... (a^{\dagger})^{\alpha_n}(b^{\dagger})_{\beta_1} 
(b^{\dagger})_{\beta_2} ... (b^{\dagger})_{\beta_m}$. The irreducible SU(3) representation states 
are the states in ${\cal{H}}^{6}_{HO}$ which are traceless in any pair of upper and lower indices 
\cite{georgi,coleman}. These SU(3) representation states are explicitly constructed in 
\cite{mm1,rmi} and given by the following polynomial of SU(3) Schwinger bosons:
\bea
\hspace{-2.4cm} |\psi\rangle ^{\alpha_1,\alpha_2,...\alpha_n}_{\beta_1,\beta_2,...,\beta_m} \equiv
\Big[ O^{\alpha_1\alpha_2...\alpha_n}_{\beta_1\beta_2...\beta_m} + L_1 \sum_{l_1=1}^n \sum_{k_1=1}^m
\delta^{\alpha_{l_1}}_{\beta_{k_1}} O^{\alpha_1\alpha_2..\alpha_{l_1-1}\alpha_{l_1 +1}..\alpha_n}_{\beta_1\beta_2..
\beta_{k_1-1} \beta_{k_1 +1} ...\beta_m}
+ L_2 \sum_{({}^{l_1,l_2}_{~=1})}^n \sum_{({}^{k_1,k_2}_{~=1})}^m
\delta^{\alpha_{l_1}\alpha_{l_2}}_{\beta_{k_1}\beta_{k_2}}
\nonumber \\
\hspace{-2.4cm}O^{\alpha_1..\alpha_{l_1-1}\alpha_{l_1+1}..\alpha_{l_2-1}\alpha_{l_2+1}..\alpha_n}_{\beta_1..
\beta_{k_1-1}\beta_{k_1+1}..\beta_{k_2-1}\beta_{k_2+1} ..\beta_m}
+ L_3 \sum_{({}^{l_1,l_2}_{l_3=1})}^n \sum_{({}^{k_1,k_2}_{k_3=1})}^m
\delta^{\alpha_{l_1}\alpha_{l_2}\alpha_{l_3}}_{\beta_{k_1}\beta_{k_2}\beta_{k_3}}
O^{\alpha_1..\alpha_{l_1-1}\alpha_{l_1+1}..\alpha_{l_2-1}\alpha_{l_2+1}.. \alpha_{l_3-1}\alpha_{l_3
+1}.. \alpha_n}_{\beta_1..\beta_{k_1-1}\beta_{k_1+1}..
\beta_{k_2-1}\beta_{k_2+1}..\beta_{k_3-1}\beta_{k_3+1} ...\beta_m} \nonumber \\
\hspace{-2.3cm} +.. + L_q \sum_{l_1..l_q=1}^n \sum_{k_1..k_q=1}^{m}
\delta^{\alpha_{l_1}\alpha_{l_2}..\alpha_{l_q}}_{\beta_{k_1}\beta_{k_2}..\beta_{k_q}}
O^{\alpha_1\alpha_2..\alpha_{l_1-1}\alpha_{l_1+1}..\alpha_{l_2-1}\alpha_{l_2+1}..
\alpha_{l_q-1}\alpha_{l_q+1}.. \alpha_n}_{\beta_1
\beta_2.. \beta_{k_1-1}\beta_{k_1+1}.. \beta_{k_2-1}\beta_{k_2+1}..\beta_{k_q-1}\beta_{k_q+1} ...
\beta_m} \Big] |0\rangle
\label{bv}
\eea
The coefficients $L_r$ are given by \cite{mm1,rmi}:
\beq
L_r \equiv {(-1)^{r} ~(a^\dagger \cdot b^\dagger )^{r} \over {(n+m+1)
(n+m)(n+m-1)...(n+m+2-r})}~,~~ r=1,2, \cdots q \equiv {\textrm{min}}(n,m), 
\label{coef}
\eeq
leading to tracelessness conditions:
\beq
\sum_{i_l,j_k=1}^{3} \delta^{\alpha_l}_{\beta_k} |\psi\rangle ^{\alpha_1,\alpha_2,...\alpha_n}_{\beta_1,
\beta_2,...,\beta_m} = 0, ~~ {\rm for ~all} ~~l=1,2...n, ~~{\rm and}~~ k=1,2...m ~.
\label{trace}
\eeq
Note that the conditions (\ref{trace}) represent a single constraint as the SU(3) irreducible states 
$|\psi\rangle ^{\alpha_1,\alpha_2,...\alpha_n}_{\beta_1, \beta_2,...,\beta_m}$ are symmetric in upper 
($\alpha$) and lower ($\beta$) indices. In fact, in the case of SU(3) the trace zero constraint is 
exactly equivalent to the vertical anti-symmetry of SU(3) Young tableau in Figure \ref{su3ytp}. 
Apart from the complicated and involved representations (\ref{bv}), the Schwinger boson construction of 
SU(3) suffers from the multiplicity problem. This problem arises because the states which differs from 
(\ref{bv}) by factors of SU(3) invariant operator $a^{\dagger}\cdot b^{\dagger}$ 
transform exactly the same way \cite{mu1,rmi}. Therefore, the infinite tower of states: 
\bea 
|\psi_{\rho} \rangle ^{\alpha_1,\alpha_2,...\alpha_n}_{\beta_1,\beta_2,...,\beta_m} \equiv 
(a^{\dagger} \cdot b^{\dagger})^{\rho} |\psi\rangle ^{\alpha_1,\alpha_2,...\alpha_n}_{\beta_1,\beta_2,...,\beta_m}~~~~, 
\label{it} 
\eea 
transform exactly like (n,m) representation in (\ref{bv}).
In the SU(2) case we do not face the multiplicity problem because the only invariant 
operator is the total number operator in (\ref{noc}) which, also being the Casimir 
simply multiplies the states in (\ref{su2irrep}) by its eigenvalue $n=2j$. 
Another equivalent and compact way to obtain and understand SU(3) irreducible representations and its multiplicity problem
is through the use of SU(3) invariant Sp(2,R) algebra \cite{mu1}. As in \cite{mu1} we 
define the following SU(3) invariant operators:
\bea 
k_{+} \equiv a^{\dagger} \cdot b^{\dagger}, ~~k_{-} \equiv a \cdot b, ~~ 
k_0 \equiv {1 \over 2} \left(N_a + N_b + 3\right).
\label{sp2ro} 
\eea
It is easy to check that they satisfy Sp(2,R) or SU(1,1) algebra:
\bea
\left[k_-,k_+\right] = 2k_0, ~~~\left[k_0,k_+\right] = k_+, ~~~\left[k_0,k_-\right] = -k_-.
\label{sp2r}
\eea
The two mutually commuting SU(3) and Sp(2,R) algebras in (\ref{su3sb}) and 
(\ref{sp2r}) were exploited in \cite{mu1} to label the infinite tower of states in 
(\ref{it}) by the additional `magnetic quantum number' of the Sp(2,R) group. 
In particular the states in (\ref{bv}) carry the lowest magnetic quantum number 
and therefore trivially satisfy: 
\bea 
k_- ~|\psi\rangle ^{\alpha_1,\alpha_2,...\alpha_n}_{\beta_1,\beta_2,...,\beta_m} 
\equiv a \cdot b ~ |\psi\rangle ^{\alpha_1,\alpha_2,...\alpha_n}_{\beta_1,\beta_2,...,\beta_m} 
= 0. 
\label{sp2rc}
\eea 
In fact, the SU(3) tracelessness constraint (\ref{trace}) and Sp(2,R) 
constraint (\ref{sp2rc}) are exactly equivalent \cite{mu1,rmi}. The constraint
(\ref{sp2rc}) also reduces the 6 dimensional harmonic oscillator Hilbert space 
${\cal{H}}^{6}_{HO}$ to the 5 dimensional SU(3) Hilbert space ${\cal{H}}^{(5)}_{SU(3)}$. 
The SU(3) Hilbert space ${\cal{H}}^{(5)}_{SU(3)}$ is generally characterized 
by $|N_a,N_b,I,M,Y\rangle$ where $N_a,N_b$ are the SU(3) Casimirs (\ref{su3cas}) fixing 
the representation. The other three quantum numbers are usually the total isospin $I$, its third component 
$M$ and hyper charge $Y$ which are the Casimirs of the chain of 
canonical subgroup SU(2) $\otimes$ U(1) $\in$ SU(3) and U(1) $\in$ SU(2) respectively. 
\subsection{The Irreducible SU(3) Schwinger Bosons}
\label{su3isb}
\begin{figure}[t]
\begin{center}
\includegraphics[width=0.5\textwidth,height=0.25\textwidth]
{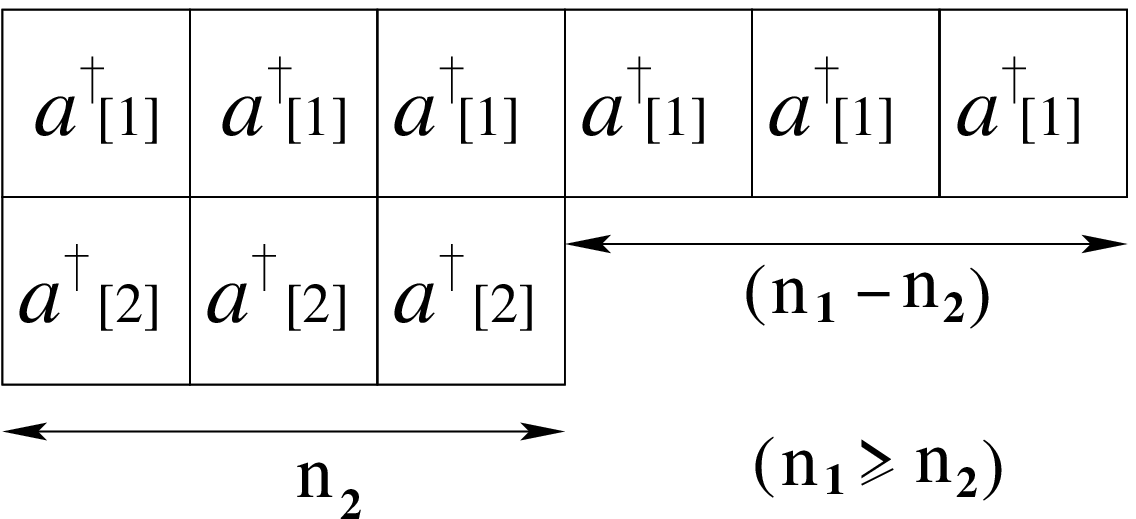}
\end{center}
\caption{SU(3) Young table in the representation $[n_1,n_2]$ with two triplets. 
The same Young tableau with a triplet and an anti-triplet in Figure \ref{su3ytp} is 
characterized by $(n,m)$ with $n=n_1-n_2$ and $m=n_2$.} 
\label{su3yt2}
\end{figure}

In recent work \cite{rmi} we have defined and constructed SU(3) irreducible creation and annihilation operators 
which directly create the SU(3) irreducible Hilbert space spanned by (\ref{bv}) from the vacuum. 
We define \cite{rmi}: 
\bea
A^{\dagger\alpha}= a^{\dagger\alpha}-\frac{1}{{N}_a+{N}_b+1}(a^\dagger \cdot b^\dagger)b^\alpha, ~~~~
A_\alpha&=& a_\alpha-b^\dagger_\alpha (a\cdot b)\frac{1}{{N}_a+{N}_b+1}\approx a_\alpha \nonumber \\ 
B^\dagger_\alpha= b^\dagger_\alpha-\frac{1}{{N}_a+{N}_b+1}(a^\dagger \cdot b^\dagger)a_\alpha, ~~~~~~
B^\alpha &=& b^\alpha-a^{\dagger{\alpha}} (a\cdot b)\frac{1}{{N}_a+{N}_b+1}\approx b^\alpha
\label{abcd1} 
\eea
In (\ref{abcd1}) $\approx$ means that that these identities are weakly satisfied. 
More explicitly, $A_{\alpha} \approx a_{\alpha}$ 
means that the actions of $A_{\alpha}$ and $a_{\alpha}$ are same on 
SU(3) irreducible Hilbert space satisfying the constraint (\ref{sp2rc}). 
It is easy to check that the irreducible Schwinger 
boson creation operators commute amongst themselves: 
\bea 
\left[A^{\dagger \alpha}, A^{\dagger \beta}\right] = 0, ~~ \left[B^{\dagger}_{\alpha}, 
B^{\dagger}_{\beta}\right] = 0, ~~ \left[A^{\dagger \alpha}, B^{\dagger}_{\beta}\right] =0. 
\label{comm} 
\eea
Hence the general SU(3) irreducible representations (\ref{bv}) which are extremely complicated 
polynomials of ordinary SU(3) Schwinger bosons can now be simply written as monomials 
of SU(3) irreducible Schwinger bosons \cite{rmi}: 
\bea
\label{modsu3irrep}
|{{\psi}}\rangle^{\alpha_1\alpha_2\ldots\alpha_n}_{\beta_1\beta_2\ldots\beta_m}
= A^{\dagger\alpha_1}A^{\dagger\alpha_2}\ldots A^{\dagger\alpha_n}
B^{\dagger}_{\beta_1}B^{\dagger}_{\beta_2}\ldots B^{\dagger}_{\beta_m}
|0\rangle ~~.
\eea 
The permutation symmetries of the upper and lower indices are inbuilt because of the commutation 
relations (\ref{comm}). The tracelessness of any mixed state $(n,m)$ is also obvious 
and immediately follows from the tracelessness of the octet state. To see this, we consider: 
\bea 
|\psi\rangle^{\gamma\alpha_2 \ldots \alpha_n}_{\gamma\beta_2\ldots \beta_m} = A^{\dagger} \cdot B^{\dagger} 
|\psi\rangle^{\alpha_2 \ldots \alpha_n}_{\beta_2 \ldots \beta_m} 
= A^{\dagger\alpha_2}\ldots A^{\dagger\alpha_n}
B^{\dagger}_{\beta_2}\ldots B^{\dagger}_{\beta_m}
|\psi\rangle^{\gamma}_{\gamma} =0 ~~~.
\label{mf}
\eea 
In (\ref{mf}), we have used the fact that all the $A^{\dagger}s$ 
and $B^{\dagger}s$ commute amongst themselves (\ref{comm}) and the octet state 
$|\psi\rangle^{\alpha}_{\beta}$ is traceless. 
We further note that: 
\bea
A \cdot B~ |\psi\rangle^{\alpha_1\alpha_2 \ldots \alpha_n}_{\beta_1\beta_2 \ldots \beta_m} \approx 
a \cdot b~ |\psi\rangle^{\alpha_1\alpha_2 \ldots \alpha_n}_{\beta_1\beta_2 \ldots \beta_m} = 0~~,\nonumber \\
A^{\dagger} \cdot B^{\dagger} ~ |\psi\rangle^{\alpha_1\alpha_2 \ldots \alpha_n}_{\beta_1\beta_2 \ldots \beta_m} = 
\sum_{\gamma=1}^{3} |\psi\rangle^{\gamma \alpha_1\alpha_2 \ldots \alpha_n}_{\gamma\beta_1\beta_2 \ldots \beta_m} = 0~~. 
\label{mf1} 
\eea
The only other SU(3) invariant operators which can be constructed out of these irreducible operators 
are $A^{\dagger} \cdot A$ and $B^{\dagger} \cdot B$ which following (\ref{sp2rc}) and (\ref{abcd1}) 
are simply the number operators $a^{\dagger} \cdot a$ and $b^{\dagger} \cdot b$ respectively. As they 
are also SU(3) Casimirs, they do not lead to any multiplicity. 
Thus this construction in terms of SU(3) irreducible Schwinger bosons also solves the 
SU(3) multiplicity problem. Further, like in SU(2) case in Figure 1, each Young tableau single (double) box $\in 3$ 
($3^*$) representation in Figure 2 corresponds to the irreducible Schwinger boson creation 
operator $A^{\dagger \alpha}$ ($B^{\dagger \alpha}$). 
In other words the defining equations of SU(3) irreducible Schwinger bosons (\ref{abcd1}) 
already take care of all the symmetries of SU(3) Young tableaues through the constraint (\ref{sp2r}). No explicit 
symmetrization or anti-symmetrization is needed. 
\noindent 
At this stage, in order to formulate SU(N) problem, it is convenient to recast the above results in terms 
of two harmonic oscillators triplets instead of a triplet and an antitriplet in (\ref{su3sb}):
\bea
Q^{\mathrm a} = a^\dagger [1]{\lambda^{\mathrm a} \over 2} a[1] + a^\dagger[2]{\lambda^{\mathrm a} \over 2} a[2] ~~.
\label{su3sbn} 
\eea
In (\ref{su3sbn}) both $a^{\dagger}[1]$ and $a^{\dagger}[2]$ transform like triplets of SU(3):
\bea
\left[Q^{\mathrm a}, a^{\dagger\alpha}[i]\right]=\left( a^\dagger[i]\frac{\lambda^{\mathrm a}}{2} \right)^\alpha ,~~~~~i=1,2.
\eea
One can get the 
anti-triplet $3^*$ representation by taking the anti-symmetric combination of the two triplets: 
$b^{\dagger}_{\alpha} = \epsilon_{\alpha\beta\gamma} a^{\dagger \beta}[1] a^{\dagger \gamma}[2]$. 
In the present representation the four SU(3) invariant operators are \cite{mosh}: 
\bea 
\hat{L}_{ij} = a^{\dagger}[i] \cdot a[j], ~~~~~i,j =1,2. 
\label{ung} 
\eea 
They satisfy U(2) algebra:
\bea 
\left[\hat{L}_{ij},\hat{L}_{kl}\right] = \delta_{jk}\hat{L}_{il} - \delta_{il}\hat{L}_{kj}.
\label{un} 
\eea 
Note that $\hat{L}_{11}$ and $\hat{L}_{22}$ are the two number operators $N_1$ and $N_2$ respectively which are also 
the two Casimirs of the SU(3) algebra in (\ref{su3sbn}). We denote their eigenvalues by $n_1$ and $n_2$ 
respectively. The corresponding SU(3) Young tableau is shown in Figure (\ref{su3yt2}). 
The advantage of the change of language from (\ref{su3sb}) to (\ref{su3sbn}) is that the constraint 
analogous to (\ref{sp2rc}) is now obvious and can be easily generalized to SU(N). The anti symmetrization 
along the $n_2$ columns of the SU(3) 
Young tableau in Figure \ref{su3yt2} immediately implies the constraint \cite{mosh}: 
\bea 
\hat{L}_{12} \equiv a^{\dagger}[1] \cdot a[2] \approx 0. 
\label{u2c} 
\eea
Note that $\hat{L}_{12}$ is an exchange operator. It replaces type [2] Schwinger boson by 
type [1] Schwinger boson. Therefore, the only solutions of the constraint (\ref{u2c}) 
are the states which are anti-symmetric in color indices along the $n_2$ columns of the 
SU(3) Young tableau in Figure \ref{su3yt2}. 
The most general forms of $A^\dagger_\alpha[1]$ and $A^\dagger_\alpha[2]$
which transform like triplets and also increase $N[1]$ and $N[2]$ by $1$ respectively 
are:
\bea
\label{nd} 
A^{\dagger\alpha}[1] &=& a^{\dagger\alpha}[1] \\
A^{\dagger\alpha}[2]&=& a^{\dagger\alpha}[2]+ F^2_1(N_1, N_2)~\hat{L}_{21}a^{\dagger\alpha}[1]~~~. \nonumber 
\eea 
The unknown coefficient $F^2_1(N_1,N_2)$ is fixed by demanding: 
\bea
(a^\dagger[1]\cdot a[2])A^{\dagger\alpha} \approx 0 => F^2_1(N_1,N_2) = -\frac{1}{N_1-N_2+2}. 
\label{dc} 
\eea 
In (\ref{dc}) we have used $\left[\hat{L}_{12},\hat{L}_{21}\right]=N_1-N_2.$
Note that the coefficient $F^2_1(N_1,N_2)$ is always well defined as the eigenvalues of $N_1$ and $N_2$ corresponding to the 
SU(3) Young tableau in Figure (\ref{su3yt2}) satisfy $n_1 \ge n_2$. The operator 
$a^{\dagger \alpha}[1]$ already commutes with the constraint operator $\hat{L}_{12}$ in (\ref{u2c}) and therefore 
remains unchanged in (\ref{nd}). In other words, the first triplet of SU(3) ISB retains the form of SU(2) Schwinger bosons. For later convenience we formally write the first equation in (\ref{nd}) as:
\bea
A^{\dagger\alpha}[1]^{SU(3)}= A^{\dagger\alpha}[1]^{SU(2)}= a^{\dagger\alpha}[1]~~.
\label{28'}
\eea
The above equation emphasizes the form invariance and iterative nature of the construction to be used later in section IV and V. 
It is easy to check that,
\bea
\Big[ A^{\dagger\alpha}[1],A^{\dagger\beta}[1] \Big]=0, ~~~ \Big[ A^{\dagger\alpha}[2],A^{\dagger\beta}[2] \Big]=0~~.
\eea
Therefore a general $(n,m)$ irrep of $SU(3)$ is obtained by these new Schwinger bosons as:
\bea
\label{modsu3irrep1}
|\psi\rangle^{(\beta_1\ldots\beta_{n_2})(\alpha_1\ldots\alpha_{n_1})}
=A^{\dagger\beta_1}[2]A^{\dagger\beta_2}[2]\ldots A^{\dagger\beta_{n_2}}[2]A^{\dagger\alpha_1}[1]
A^{\dagger\alpha_2}[1]\ldots A^{\dagger\alpha_{n_1}}[1]
|0\rangle ~~~.
\label{sm}
\eea 
The simple monomial construction (\ref{modsu3irrep1}) is equivalent to the SU(3) young tableaux with appropriate symmetries. 
To see this we construct the simplest mixed octet representation with $n_1=2$ and $n_2=1$ in (\ref{modsu3irrep1}): 
\bea
\label{(1,1)}
&& \hspace{1cm} |\psi\rangle^{(\beta)(\alpha_1\alpha_2)} = 
A^{\dagger\beta}[2]A^{\dagger\alpha_1}[1]A^{\dagger\alpha_2}[1] |0\rangle 
= A^{\dagger\beta}[2]A^{\dagger\alpha_2}[1]A^{\dagger\alpha_1}[1] |0\rangle \\
&& =\frac{1}{3} \Big\{\left(a^{\dagger\beta}[2]a^{\dagger\alpha_1}[1]-a^{\dagger\alpha_1}[2]a^{\dagger\beta}[1] 
\right)a^{\dagger\alpha_2}[1] + \left( a^{\dagger\beta}[2]a^{\dagger\alpha_2}[1]
-a^{\dagger\alpha_2}[2]a^{\dagger\beta}[1]\right)
a^{\dagger\alpha_1}[1]\Big\}|0\rangle ~~~. \nonumber 
\eea
The expression in (\ref{(1,1)}) is first antisymmetrized amongst the column indices $\alpha_1$ and $\beta$, then symmetrized amongst the row indices $\alpha_1 ~\& ~\alpha_2$. Therefore it has the symmetries of SU(3) Young tableaux in figure 3.
This simplest but non-trivial example illustrates the usefulness of the procedure to 
compute SU(3) representations in terms of SU(3) irreducible Schwinger bosons. {\it The simple 
monomial in (\ref{sm}) captures all the symmetries of the SU(3) Young tableau diagram in 
Figure (\ref{su3yt2}).} Destruction operators corresponding to (\ref{nd}) can be constructed 
by canonical conjugation of $A^{\dagger}[1]$ and $A^{\dagger}[2]$. However, these operators 
will not commute with the constraint (\ref{u2c}) and will take us out of ${\cal H}^5_{SU(3)}$. 
On the other hand, we can also construct the SU(3) irreducible 
destruction operators weakly commuting with $\hat{L}_{12}$. The construction in exactly 
same as that of $A^{\dagger\alpha}[2]$ and one obtains: 
\bea
A_\alpha[1]&=& a_\alpha[1]+\frac{1}{N_1-N_2+2} ~ \hat{L}_{21}~a_\alpha[2] ~~,\nonumber \\
A_\alpha[2]&=& a_\alpha[2]~~.
\eea
The irreducible Schwinger boson representations are free from any kind of multiplicity 
problems as all the non-trivial SU(3) invariant operators are weakly zero: 
\bea
\label{gr1} 
A^\dagger[1]\cdot A[2] & \equiv & a^\dagger[1] \cdot a[2] \equiv \hat{L}_{12} \approx 0, \\ 
A^\dagger[2]\cdot A[1] & = & 
\left(a^{\dagger\alpha}[2]-\frac{1}{N_1-N_2+2}~ \hat{L}_{21}~a^{\dagger\alpha}[1]\right)
\left(a_\alpha[1]+\frac{1}{N_1-N_2+2} ~ \hat{L}_{21}~a_\alpha[2]\right) \nonumber \\
\label{gr2} 
& = & - \frac{1}{\left(N_1-N_2+2\right) \left(N_1-N_2+3\right)} ~\left(\hat{L}_{21}\right)^2 ~\hat{L}_{12} 
\approx 0, \\ 
A[1] \cdot A^\dagger[2] & = & 
\left(a_\alpha[1]+\frac{1}{N_1-N_2+2} ~ \hat{L}_{21}~a_\alpha[2]\right) 
\left(a^{\dagger\alpha}[2]-\frac{1}{N_1-N_2+2}~ \hat{L}_{21}~a^{\dagger\alpha}[1]\right) \nonumber \\
\label{gr3} 
& = & - \frac{1}{\left(N_1-N_2+2\right) \left(N_1-N_2+3\right)} ~\left(\hat{L}_{21}\right)^2 ~\hat{L}_{12} \approx 0. 
\eea
Note that in calculating $A^\dagger[2]\cdot A[1]$ and $A[1]\cdot A^\dagger[2]$ in (\ref{gr2}) and (\ref{gr3}) 
respectively all the linear terms in $\hat{L}_{21}$ cancel out exactly and the quadratic term 
$\left(\hat{L}_{21}\right)^2$ is proportional to the constraint $\hat{L}_{12}\approx 0$ in (\ref{u2c}). Note that the total ISB number operators are trivial invariant operators as:
\bea
A^\dagger[1]\cdot A[1]\approx a^\dagger[1] \cdot a[1] ~~~~\& ~~~~ A^\dagger[2]\cdot A[2]\approx a^\dagger[2] \cdot a[2]~~~.
\eea 
As mentioned in the introduction at this stage it is illustrative to give explicit construction of 
SU(4) irreducible Schwinger bosons before dealing with SU(N) in section (\ref{sun}). 
\section{SU(4) irreducible representations}
\label{su4}
\begin{figure}[t]
\begin{center}
\includegraphics[width=0.6\textwidth,height=0.3\textwidth]
{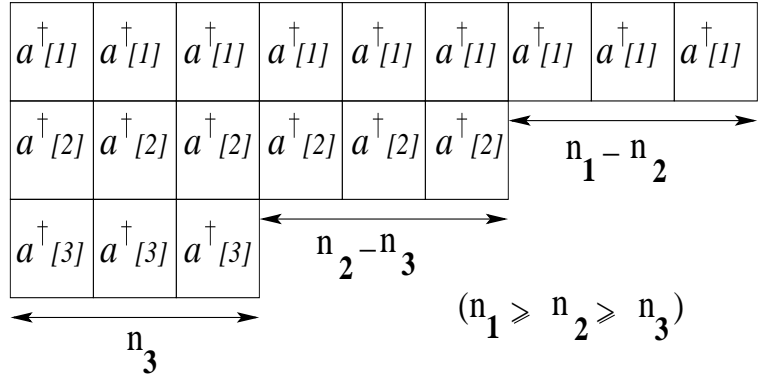}
\end{center}
\caption{SU(4) Young tableau for the representation $[n_1,n_2,n_3].$} 
\label{su4yt}
\end{figure}

The rank of SU(4) group is 3. Therefore, as shown in Figure (\ref{su4yt}), we need three 4-plets 
$a^{\dagger \alpha}[i], i=1,2,3$ to construct any irreducible representation of SU(4). The SU(4) 
generators in terms of these Schwinger bosons are: 
\bea 
Q^{\mathrm a} = a^\dagger [1]{\Lambda^{\mathrm a} \over 2} a[1] 
+ a^\dagger[2]{\Lambda^{\mathrm a} \over 2} a[2] 
+ a^\dagger[3]{\Lambda^{\mathrm a} \over 2} a[3], ~~~~~ {\mathrm a}=1,2, \cdots ,15. 
\label{su4sb}
\eea
In (\ref{su4sb}) $\Lambda^{\mathrm a}$ are the $4 \times 4$ representations of SU(4) Lie algebra. 
The $12$ harmonic oscillators in (\ref{su4sb}) create a 12 dimensional Hilbert space 
${\cal{H}}^{12}_{HO}$. The SU(4) invariant group is now U(3) consisting of 9 generators 
$\hat{L}_{ij}$ in (\ref{ung}) and (\ref{un}) with $i,j=1,2,3$. Like in SU(3) case, 
we obtain the vertical anti-symmetry of SU(4) Young tableau in Figure (\ref{su4yt}) 
by demanding \cite{mosh}:
\bea
\label{c1}
\hat{L}_{12}= a^\dagger[1]\cdot a[2]&\approx & 0 \\
\label{c2}
\hat{L}_{13}=a^\dagger[1]\cdot a[3]&\approx & 0 \\
\label{c3} 
\hat{L}_{23}=a^\dagger[2]\cdot a[3]&\approx & 0. 
\eea
The three constraints (\ref{c1}),(\ref{c2}) $\&$ (\ref{c3}) impose the vertical anti symmetries 
in the first-second, first-third $\&$ second-third rows respectively of the 
SU(4) Young tableau shown in Figure {\ref{su4yt}. 
Therefore the null space of $[\hat{L}_{12},\hat{L}_{13},\hat{L}_{23}]$ 
within ${\cal{H}}^{12}_{HO}$ is the space of SU(4) irreducible 
representation. This space is $9 (=12-3) $ dimensional and will be denoted by ${\cal H}^9_{SU(4)}$. Amongst the 
nine quantum numbers labeling ${\cal H}^9_{SU(4)}$, three are the eigenvalues of the three SU(4) 
Casimir number operators: $\hat{L}_{ii} = a^{\dagger}[i] \cdot a[i]$ with $i=1,2,3$. 
The remaining 6 magnetic quantum numbers characterizing a specific state within the above representation 
are usually chosen to be the eigenvalues of the Casimirs of the canonical subgroup 
chain: SU(3) $\otimes$ U(1) $\in$ SU(4); SU(2) $\otimes$ U(1) $\in$ SU(3) and U(1) $\in$ SU(2). 
Thus, like in SU(3) case, the dimension of SU(4) representation space computed through canonical way 
matches with the constraint analysis above (also see \cite{mosh}).
We now come to the explicit construction. We would like to construct three types of SU(4) 
irreducible Schwinger bosons $A^{\dagger}[i]$ with $i=1,2,3$ so that the monomial states: 
\bea
\hspace{-0.8cm} {\Big|\left\{\alpha _1\ldots\alpha_{n_1}\right\} 
\left\{\beta_1\ldots\beta_{n_2}\right\} 
\left\{\gamma _1\ldots\gamma_{n_3}\right\} \Big \rangle}_{n_1 \ge n_2 \ge n_3} 
& \equiv & {\Big(A^{\dagger \gamma_1}[3] \cdots A^{\dagger \gamma_{n_3}}[3]\Big)} 
{\Big(A^{\dagger \beta_1}[2] \cdots A^{\dagger \beta_{n_2}}[2]\Big)} \nonumber \\ 
&& {\Big(A^{\dagger \alpha_1}[1] \cdots A^{\dagger \alpha_{n_1}}[1]\Big)} |0\rangle 
\label{su4ir} 
\eea 
carry the symmetries of SU(4) Young tableau shown in Figure (\ref{su4yt}). Note that
the ordering is important in (\ref{su4ir}). 
Like in SU(3) case, 
the irreducible Schwinger bosons that increase $N_1,~N_2,~N_3$ by one respectively are 
constructed as:
\bea
\label{SU4A1}
A^{\dagger\alpha}[1]&=& a^{\dagger\alpha}[1] \\
\nonumber\label{SU4A2}
A^{\dagger\alpha}[2] &=& a^{\dagger\alpha}[2] + F^2_1(N_1,N_2,N_3) ~ \hat{L}_{21} ~a^{\dagger\alpha}[1] \\
\label{SU4A3}
A^{\dagger\alpha}[3] &=& a^{\dagger\alpha}[3]+ F^3_2(N_1,N_2,N_3) ~ \hat{L}_{32} ~ a^{\dagger\alpha}[2] + 
F^3_1(N_1,N_2,N_3) ~ \hat{L}_{31} ~ a^{\dagger\alpha}[1]\nonumber \\
& + & F^3_{21} (N_1,N_2,N_3) \Big( \hat{L}_{32} \hat{L}_{21} \Big) a^{\dagger\alpha}[1]~~~.\nonumber
\eea
Note that the three constraints are not independent: 
$$ \left[\hat{L}_{12},\hat{L}_{23}\right]=\hat{L}_{13}.$$
Hence implementation of the two constraints (\ref{c1}) and (\ref{c3}) should enable 
us to compute all the four structure functions $F$. 
It is clear from (\ref{SU4A1}) that $A^{\dagger\alpha}[1]$ commutes with all the constraint given in 
(\ref{c1},\ref{c2},\ref{c3}). The form of $A^{\dagger\alpha}[2]$ in (\ref{SU4A2}) is exactly same as 
in the case of SU(3) in (\ref{nd}) except that $\alpha$ runs from $1$ to $4$ and the number operators 
correspond to that of SU(4). Thus following the same method with the constraint (\ref{c1}) we obtain the same solution (\ref{dc}) 
for $F^2_1(N_1,N_2,N_3)$:
\bea
F^2_1(N_1,N_2,N_3)=F^2_1(N_1,N_2) = -\frac{1}{N_1-N_2+2}~~~,
\label{f21} 
\eea
leading to,
\bea
A^{\dagger\alpha}[2]&=& a^{\dagger\alpha}[2]-\frac{1}{N_1-N_2+2}~ \hat{L}_{21} ~a^{\dagger\alpha}[1]~~~.
\label{su4A2}
\eea
Note that the other two constraints $\hat{L}_{13}$ and $\hat{L}_{23}$ are already satisfied 
by $A^{\dagger\alpha}[2]$ in (\ref{su4ir}), i.e.: 
\bea
\hat{L}_{13} ~ A^{\dagger\alpha}[2] & \approx & \Big[\hat{L}_{13},A^{\dagger\alpha}[2] \Big]
= \frac{1}{N_1-N_2+1}~ \hat{L}_{23} ~a^{\dagger\alpha}[1]\approx 0 \nonumber \\
\hat{L}_{23} ~ A^{\dagger\alpha}[2] & \approx& 0. 
\eea
Imposing (\ref{c1}) and (\ref{c3}) and after some algebra we get: 
\bea
\hspace{-0.3cm} F^3_2 = -\frac{1}{(N_2-N_3+2)}; ~F^3_1 = -\frac{1}{(N_1-N_3+3)}; 
F^3_{21} = \frac{1}{(N_2-N_3+2)(N_1-N_3+3)} \equiv F^3_2~F^3_1. 
\label{F3}
\eea
\noindent
Like in SU(3) case, the construction of the SU(4) irreducible destruction operators is similar and one obtains: 
\bea
A_\alpha[3]&=& a_\alpha[3] \\ 
A_\alpha[2]&=& a_\alpha[2]+\frac{1}{N_2-N_3+2}~\hat{L}^{\dagger}_{23}~ a_\alpha[3] \nonumber \\
A_\alpha[1]&=& a_\alpha[1] +\frac{1}{N_1-N_2+2} ~\hat{L}^{\dagger}_{12}~ a_\alpha[2] 
+\frac{1}{N_1-N_3+3} ~\hat{L}^{\dagger}_{13} a_\alpha[3]
\nonumber\\
&& +\frac{1}{(N_1-N_2+2)(N_1-N_3+3)} ~\hat{L}^{\dagger}_{12} \hat{L}^{\dagger}_{23}~a_\alpha[3]. \nonumber
\eea 
One can easily check the commutation relations: 
\bea 
\left[A^{\dagger\alpha}[i],A^{\dagger\beta}[i]\right] = 0;~~~ i=1,2,3. 
\label{commrel} 
\eea 
In fact, the above identity is trivial for i =1 as $A^{\dagger \alpha}[1] \equiv a^{\dagger \alpha}[1]$. 
Thus the antisymmetrizations amongst the SU(4) Young tableau column indices in 
(Figure \ref{su4yt}) are implemented by imposing constraints (\ref{c1}), (\ref{c2}) and (\ref{c3}) on 
the SU(4) irreducible Schwinger bosons. 
Having antisymmetrized this way, the commutation relations (\ref{commrel}) ensure the horizontal 
permutation symmetries amongst the indices belong to each of the three rows of Figure \ref{su4yt}. 
As a result the resultant ISB monomial states in (\ref{su4ir}) carry all the symmetries of SU(4) 
Young tableau. Further these representations are also multiplicity free as: 
\bea
A[1]\cdot A^\dagger[2]&\approx & 0 ~~,~~~~~~~ A[2]\cdot A^\dagger[1]\approx 0 \\ 
A[1]\cdot A^\dagger[3]&\approx & 0 ~~,~~~~~~~ A[3]\cdot A^\dagger[1]\approx 0 \\ 
A[2]\cdot A^\dagger[3]&\approx & 0 ~~,~~~~~~~ A[3]\cdot A^\dagger[2]\approx 0. 
\eea
The above SU(4) results are analogues of SU(3) results in (\ref{gr1}), (\ref{gr2}) and (\ref{gr3}). 
\noindent An alternative irreducible Schwinger boson construction procedure is to exploit the iterative nature of the solutions. 
In the present SU(4) case we note that $A^{\dagger \alpha}[1]$ and $A^{\dagger \alpha}[2]$ 
retain the same form as SU(3) and only $A^{\dagger \alpha}[3]$ has to be constructed to satisfy 
all the three constraints. In the construction of $A^{\dagger \alpha}[3]$ also one can use 
SU(3) ISB so that the first fundamental constraint ($\hat{L}_{12} \approx 0$) becomes redundant and only 
the last constraint ($\hat{L}_{23} \approx 0$) has to be implemented by hand. 
Like in SU(3) case (see eqn. (\ref{28'}), we stress on this form invariance by rewriting the three equations in (\ref{SU4A1}) as: 
\bea
\label{su4m}
A^{\dagger \alpha}[1]^{[SU(4)]}&=& A^{\dagger \alpha}[1]^{[SU(3)]}= A^{\dagger \alpha}[1]^{[SU(2)]}\\
A^{\dagger \alpha}[2]^{[SU(4)]}&=& A^{\dagger \alpha}[2]^{[SU(3)]}\nonumber\\
A^{\dagger \alpha}[3]^{[SU(4)]}&=& a^{\dagger \alpha}[3]+G^3_2 \left(a^\dagger[3]\cdot A[2]^{[SU(3)]}\right)A^{\dagger \alpha}[2]^{[SU(3)]}\nonumber\\
&& +G^3_1 \left(a^\dagger[3]\cdot A[1]^{[SU(3)]}\right)A^{\dagger \alpha}[1]^{[SU(3)]}~~~. \nonumber 
\eea
Note that all the constraints in (\ref{c1}, \ref{c2} and \ref{c3}) are trivially satisfied by $A^{\dagger \alpha}[1]^{[SU(4)]}$ and $A^{\dagger \alpha}[2]^{[SU(4)]}$ by construction.
Now $\hat{L}_{23} = a^\dagger[2]\cdot a [3]\approx 0$ can be solved for $G^3_2$ and $G^3_1$ as: 
\bea
\label{su4f}
G^3_2 = -\frac{1}{N_2-N_3+2}~~~~\& ~~~~ G^3_1 = -\frac{N_1-N_2+2}{(N_1-N_2+1)(N_1-N_3+3)}~~.
\eea
One can check explicitly that with these coefficients the construction in (\ref{su4m}), with the coefficients in (\ref{f21}) and (\ref{F3}), is exactly same as 
(\ref{SU4A1}). We will use this iterative construction for SU(N) in the next section.
\section{SU(N) Irreducible Schwinger Bosons}
\label{sun}
The fundamental constituents required to construct any arbitrary irrep of SU(N) are $N-1$ independent Schwinger boson $N$-plets given by $a^{\dagger\alpha}[1]$, $a^{\dagger\alpha}[2]$, $a^{\dagger\alpha}[3]$,...,$a^{\dagger\alpha}[N-1]$, with $\alpha =1,2,3,..,N$ since the rank of the group $SU(N)$ is $N-1$. The SU(N) generators in terms of these Schwinger bosons are:
\bea
\label{sungen}
Q^{\mathrm a}= \sum _{i=1}^{N-1}a^\dagger [i]\,\frac{\Lambda^{\mathrm a}}{2}\,a[i] ~~~,
\eea
where, $\Lambda^{\mathrm a} $'s are the generalization of Gell-Mann matrices for SU(N). The $N(N-1)$ Harmonic 
oscillators present in (\ref{sungen}) creates a $N(N-1)$ dimensional Hilbert space $\mathcal H^{N(N-1)}_{\mathrm 
{HO}}$. There are $(N-1)$ Casimirs associated with SU(N) group. 
The SU(N) invariant group is now U(N-1) with $(N-1)^2$ generators given by $\hat L_{ij},~~i,j=1,2,..,N-1$.
In the representation (\ref{sungen}) the $(N-1)$ Casimirs are the number operators 
$\hat{L}_{ii} \equiv N[i] = a^{\dagger}[i] \cdot a[i]$ with $i=1,2,..,N-1$. Their eigenvalues, 
specifying a particular representation are denoted by: 
$(n_1,n_2, \cdots \cdots ,n_{N-1})$ respectively. 
For arbitrary SU(N) we obtain the vertical antisymmetry of an Young tableaux by imposing the constraints \cite{mosh}: 
\bea
\hat L_{ij}=a^\dagger[i]\cdot a[j]\approx 0, ~~~~ \mbox{for $i<j$ and $i,j=1,2,..N-1$ for SU(N).}
\label{unc} 
\eea 
\begin{figure}[t]
\begin{center}
\includegraphics[width=0.8\textwidth,height=0.35\textwidth]
{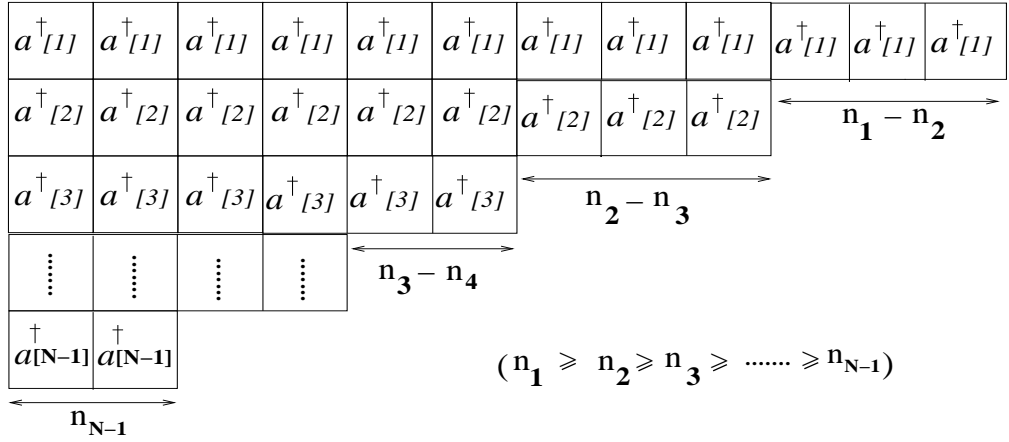}
\end{center}
\caption{SU(N) Young tableau for the representation $[n_1,n_2, \cdots n_{N-1}].$} 
\label{sunyt}
\end{figure}
Therefore the null space of $\hat L_{ij}$ for $i<j$ within $\mathcal H^{N(N-1)}_{\mathrm 
{HO}}$ is the space of SU(N) irreducible representations.
There are $\frac{1}{2}(N-1)(N-2)$ constraints in (\ref{unc}). Therefore the dimension of the 
Hilbert space containing SU(N) representations in Figure \ref{sunyt} is 
$N(N-1)-\frac{1}{2}(N-1)(N-2)=\frac{1}{2}(N-1)(N+2)$. 
The remaining $ \frac{1}{2} N (N-1) \left(=\frac{1}{2}(N-1)(N+2) - (N-1)\right)$ 
SU(N) 'magnetic quantum numbers' 
specify a particular state within the above representation. These magnetic quantum numbers 
are usually taken as the eigenvalues of the $\frac{1}{2} N (N-1)$ Casimirs of the canonical subgroup 
chain: $$\mbox{SU(N-1)} \otimes \mbox{U(1)} \in \mbox{SU(N)}, \mbox{ SU(N-2)} \otimes \mbox{U(1)} \in \mbox{SU(N-1)}, \cdots\cdots ,\mbox{U(1)} \in \mbox{SU(2)}.$$ 
We now come to the explicit construction. 
We would like to construct $N-1$ types of SU(N) 
irreducible Schwinger bosons $A^{\dagger}[i]$ with $i=1,2,..,N-1$ so that the states: 
\bea
\label{sunir} 
&&\Bigg|\left\{\alpha^{[1]} _1\alpha^{[1]} _2\ldots\ldots\alpha^{[1]}_{n_1}\right\}
\left\{\alpha^{[2]}_1\alpha^{[2]}_2 \ldots\ldots \alpha^{[2]}_{n_2}\right\} \cdots \cdots 
\left\{\alpha^{[N-1]}_1\alpha^{[N-1]}_2\ldots\alpha^{[N-1]} _{n_{N-1}}\right\} \Bigg\rangle \equiv 
\\ 
&& 
{\left\{A^{\dagger \alpha^{[N-1]}_1}[N-1] \cdots A^{\dagger \alpha^{[N-1]}_{n_{N-1}}}[N-1]\right\}}
\cdots \cdots 
{\left\{A^{\dagger \alpha^{[2]}_1}[2] \cdots A^{\dagger \alpha^{[2]}
_{n_2}}[2]\right\}} 
{\left\{A^{\dagger \alpha^{[1]}_1}[1] \cdots A^{\dagger \alpha^{[1]}_{n_1}}[1]\right\}}
|0\rangle \hspace{1.5cm} \nonumber 
\eea 
carry all the symmetries of SU(N) Young tableau. 
As discussed in the SU(4) section, $A^{\dagger}_{\alpha}[k], ~k=1,2,....,N-2$ for SU(N) have exactly the 
same form as SU(N-1). We only need to construct $A^{\dagger}_{\alpha}[N-1]$. We construct 
this in terms of SU(N-1) ISB so that 
we have to implement only the last fundamental constraint: 
$$\hat L_{(N-2)(N-1)}=a^\dagger[N-2]\cdot a[N-1] \approx 0.$$
Thus the SU(N) irreducible Schwinger bosons are given by,
\bea
\label{xyz} 
A^{\dagger \alpha}[1]^{SU(N)}&=& A^{\dagger \alpha}[1]^{SU(N-1)}= A^{\dagger \alpha}[1]^{SU(N-2)}=\cdots\cdots\cdots= A^{\dagger \alpha}[1]^{SU(2)} \nonumber \\
A^{\dagger \alpha}[2]^{SU(N)}&=& A^{\dagger \alpha}[2]^{SU(N-1)}=A^{\dagger \alpha}[2]^{SU(N-2)}=\cdots\cdots = A^{\dagger \alpha}[2]^{SU(3)} \nonumber \\
A^{\dagger \alpha}[3]^{SU(N)}&=& A^{\dagger \alpha}[3]^{SU(N-1)}=A^{\dagger \alpha}[3]^{SU(N-2)}=\cdots = A^{\dagger \alpha}[3]^{SU(4)} \nonumber \\
&\vdots & \\
A^{\dagger \alpha}[N-2]^{SU(N)}&=& A^{\dagger \alpha}[N-2]^{SU(N-1)}, \nonumber \\
A^{\dagger\alpha}[N-1]^{SU(N)}&=& a^{\dagger\alpha}[N-1]+
\sum_{i=1}^{N-2}G^{N-1}_i\left(a^\dagger[N-1]\cdot A[i]^{SU(N-1)}\right), 
A^{\dagger\alpha}[i]^{SU(N-1)}. \nonumber 
\eea
Note that in (\ref{xyz}) $\alpha =1,2, \cdots ,N$ and all the number operators are of n-plets of SU(N). 
We again emphasize that the first $(N-3)$ fundamental constraints $$\hat L_{i,i+1} = 
a^\dagger[i]\cdot a[i+1] \approx 0, ~~ i=1,2, \cdots N-3 $$ are already satisfied by the SU(N) ISB in (\ref{xyz}). 
Unfolding all the ISB constructions from SU(N-1) to SU(2) one obtains the most general form of 
$k^{th}$ irreducible Schwinger bosons: 
\bea
A^{\dagger\alpha}[k] &=& a^{\dagger\alpha}[k]+\sum_{r=1}^{k-1}
\sum_{\{i_1,..,i_r\}=1}^{{k-1}_{\prime}} ~F^k_{i_1}~F^k_{i_2} \cdots F^k_{i_r} \hat L_{ki_1}~ \hat L_{i_1i_2}
\ldots\hat L_{i_{r-1}i_r}a^{\dagger\alpha}[i_r]. 
\label{sunirsb} 
\eea
In (\ref{sunirsb}) $k=1,2, \cdots (N-1)$ and the prime over the second summation ($\sum^{\prime}$) implies 
that the ordering $k>i_1>i_2>...>i_r$ has to be maintained. 
To find out the general form of $F^k_i(N_1,..,N_{N-1})$ it is sufficient to apply only the consecutive 
$(k-1)$ fundamental constraints $\hat L_{p(p+1)}\approx 0$, where $p=1,2,..., k-1$ as all other 
constraints can be obtained as the commutators of these fundamental constraints, e.g.: 
$\hat{L}_{13} = \left[\hat{L}_{12},\hat{L}_{23}\right], \hat{L}_{14} = \left[\hat{L}_{13},\hat{L}_{34}\right]
= \left[\left[\hat{L}_{12},\hat{L}_{23}\right],\hat{L}_{34}\right]$ etc.. 
It is quite straightforward to show that the constraint $\hat L_{(N-1)N}\approx 0$ gives 
\bea
\label{Hk-1}
F^k_{k-1}(N_1,..,N_{N-1})=F^k_{k-1}(N_k,N_{k-1}) =-\frac{1}{N_{k-1}-N_k+2}~~~.
\eea
After some algebra, the general constraints $a^\dagger[i]\cdot a[i+1]\approx 0$ gives the recurrence relation:
\bea
F^k_{p}(N_1,..,N_{N-1})=\frac{F^k_{p+1}(N_1,..,N_{N-1})}{1-(N_p-N_{p+1}+1)F^k_{p+1}(N_1,..,N_{N-1})}~~~. 
\label{x1} 
\eea
The solution of (\ref{x1}) with (\ref{Hk-1}) as the boundary condition is: 
\bea
F^k_i=-\frac{1}{N_i-N_k+1+k-i}~~~.
\eea
Note that these SU(N) solutions for $F^{k}_i$ reduce to (\ref{dc}) and (\ref{F3}) for N=3 and 4 respectively. 
Similarly all the $N-1$ fundamental irreducible annihilation operators for SU(N) can also be constructed using the 
irreducible creation and annihilation operators for SU(N-1). The general $k^{th}$ annihilation operator for SU(N) 
is given by,
\bea
A_\alpha[k]^{SU(N)}= a_\alpha[k]+\sum_{i=k+1}^{N-1}G'^i_K\left( a[K]\cdot A^\dagger[i]^{SU(N-1)} \right)A_\alpha[i]^{SU(N-1)}~~~,
\eea
or equivalently,
\bea
A_\alpha[k] \equiv a_\alpha[k]+\sum _{r=1}^{N-1} \sum_{\{i_1,i_2,..,i_r= k+1\}}^{{N-1}_{\prime}}~ 
H^{i_1}_k H^{i_2}_k\ldots H^{i_r}_k \hat L_{i_1k}\hat L_{i_2i_1}\ldots \hat L_{i_ri_{r-1}}a_\alpha[i_r]. 
\label{abcd2} 
\eea
In (\ref{abcd2}) $k=1,2, \cdots (N-1)$ and the prime over the second summation ($\sum^{\prime}$) implies 
that the ordering $k<i_1<i_2<...<i_r<N-1$ has to be maintained. The similar algebra as done 
for creation operators gives:
\bea
H^i_k= \frac{1}{N_i-N_k+1+k-i} \equiv -F^k_i~~~.
\eea
The Hilbert space created by SU(N) irreducible Schwinger bosons contains all SU(N) representations and 
every representation appears once as: 
$$ A^\dagger[i]\cdot A[j]\approx 0,~~~~\forall~~ i\ne j~~~.$$
The only remaining SU(N) invariant operators in terms of SU(N) ISB are 
$A^\dagger[i]\cdot A[i], ~ i=1,2, \cdots ,(N-1).$ These operators, being weakly 
related to the SU(N) number operator Casimirs, do not lead to multiplicity. 
\section{Summary and Discussions}
We conclude that SU(N) representations constructed in terms of SU(N) ISB are 
complete as well as economical. Further, like in SU(2) case, all representations 
are monomials of SU(N) ISB. This is because the SU(N) ISB are defined and constructed 
such that they carry the symmetries of SU(N) Young tableaues making explicit 
symmetrization, antisymmetrizations redundant. Thus the SU(N) irreducible Schwinger bosons 
($N \ge 3$) provide a model for SU(N) just like SU(2) Schwinger bosons provide a model 
for SU(2). At this stage, following the discussion in \cite{sharat2}, 
it is interesting to mention the parallels with quantization of gauge theories. In gauge 
theories there are many spurious 
gauge degrees of freedom. Therefore, all states connected by gauge 
transformations represent a single physical state. This multiplicity is removed 
by imposing the Gauss law constraint on the physical Hilbert space. In the case of 
pure electrodynamics the Gauss law constraint is: 
\bea 
\bigtriangledown \cdot E ~|\Psi\rangle_{\textrm{physical}} = 0~~~~ {\textrm{or}} ~~~~~~ 
\bigtriangledown \cdot E \approx 0. 
\label{gl}
\eea 
In (\ref{gl}) $\bigtriangledown \cdot E$ represents the divergence of electric field. 
This is analogous to the constraints (\ref{sp2rc},\ref{unc}) which can be interpreted as 
`group theory Gauss law' constraints in ${\cal H}^{N(N-1)}_{HO}$. The representation 
redundancy in ${\cal H}^{N(N-1)}_{HO}$ is generated by the generators of invariant $U(N-1)$ 
group.
\noindent In the case of SU(3), discussed in detail in \cite{mu1} the six dimensional harmonic oscillator Hilbert space ${\cal H}^{6}_{HO}$ was 
completely spanned by vectors labeled with the 6 quantum numbers belonging to $SU(3) \otimes Sp(2,R)$ group. 
Similarly, the harmonic oscillator Hilbert space ${\cal H}^{N(N-1)}_{HO}$ can also be completely 
spanned by vectors labeled by the $N(N-1)$ quantum numbers belonging to $SU(N) \otimes U(N-1)$ or equivalently 
$SU(N) \otimes SU(N-1) \otimes U(1)$. The work in this direction is in progress and will be reported elsewhere. 
\noindent
An important application of SU(N) ISB is in lattice gauge theories \cite{rmi2}. 
The SU(N) ISB enable us to remove the redundant gauge as well as loop degrees of freedom from lattice 
gauge theories \cite{lgt,rmi2} leading to a formulation in terms of loops and strings without any spurious gauge or loop degrees of freedom. In fact, this has been the starting point and the motivation behind 
the present work and the work in \cite{rmi,rmi2}. 


\begin{thebibliography}{99}
\bibitem{schwinger} J. Schwinger U.S Atomic Energy Commission Report NYO-3071, 
1952 or D. Mattis, {\it The Theory of Magnetism} (Harper and Row, 1982). 
\bibitem{np} Marshalek E R, Phys. Rev. C {\bf 11}, (1975) 1426.
\bibitem{scs} Auerbach A and Arovas D P, Phys. Rev. Lett. {\bf 61}, 617 (1988). \\
Auerbach A 1994 Interacting Electrons and Quantum Magnetism (Berlin: Springer).\\
Sachdev S and Read N, Nucl. Phys. B {\bf 316} 609 (1989).
\bibitem{susy} Gunaydin M and Saclioglu C, Commun. Math. Phys. {\bf 87}, 159 (1982).\\
Gunaydin M and Saclioglu C, Phys. Lett. B {\bf 108}, 180 (1982).
\bibitem{lgt} Manu Mathur, Nucl.\ Phys.\ B {\bf 779}, 32 (2007). \\
Manu Mathur, Phys.\ Lett.\ B {\bf 640}, 292 (2006).\\ 
Manu Mathur, J.\ Phys.\ A {\bf 38}, 10015 (2005). \\
Chandrasekharan S and Wiese U J, Nucl. Phys. B {\bf 492}, 455 (1997).
\bibitem{lqg} Florian Girelli, Etera R. Livine, Class.Quant.Grav. {\bf 22}, 3295 (2005). \\
N. D. Hari Dass, Manu Mathur, Class.Quant.Grav. {\bf 24}, 2179 (2007). 
\bibitem{mosh} M. Moshinsky, Rev. Mod. Phys. {\bf 34}, 813 (1962); J. Math. Phys. {\bf 4}, 1128 (1963).
\bibitem{gel} I.N. Bernstein, I.M. Gelfand, S.I. Gelfand, Funct. Anal. Appl. {\bf 9}, 322 (1975).\\ 
I.M. Gelfand and A.V. Zelevinskii, Funct. Anal. Appl. {\bf 18}, 183 (1984). \\
Bemstein I N, Gelfand I M and Gclfand S I 1976 Proc Pefrovskij Sem. 2 3
Reprinted in Gelfand I 1988 Collecfed Works VoL 1 (Berlin: Springer) p 464. 
\bibitem{apw} 
J. J. De Swart, Rev. Mod. Phys. {\bf 35}, 916 (1963).\\ 
J. D. Louck, Am. J. Phys. {\bf 38}, 3 (1970).\\
C. Itzykson, Rev. Mod. Phys. {\bf 38}, 95 (1966). \\ 
Arisaka N, Prog. Theor. Phys. {\bf 47}, 1758 (1972). \\
N. Mukunda and L. K. Pandit, J. Math. Phys. {\bf 6}, 746 (1965).\\
M. Resnikoff, J. Math. Phys. {\bf 8}, 63 (1967).\\
P. Jasselette, Nucl. Phys. B {\bf 1}, 521 (1967); ibid 529. \\
P. Jasselette, J. Phys. A: Math. Gen., {\bf 13}, 2261, (1980).\\ 
Biedenhm L C, J. Math. Phys. {\bf 4} 436 (1963). \\
R.~Anishetty, H.~Gopalkrishna Gadiyar, M.~Mathur and H.~S.~Sharatchandra,
Phys.\ Lett.\ B {\bf 271}, 391 (1991). 
\bibitem{sharat} J. S.Prakash and H. S. Sharatchandra,
J.\ Math.\ Phys.\ {\bf 37}, 6530 (1996) and references cited therein. 
\bibitem{sharat2} J. S. Prakash and H. S. Sharatchandra, J.\ Phys.\ A {\bf 26}, 1625 (1993).
\bibitem{bied} Biedenharn, L.C., Flath, D.E., 
Commun. Math. Phys. {\bf 93} , 143 (1984).\\ A. J. Bracken, Commun. Math. Phys. {\bf 94}, 371 (1984). \\
A J Bracken and J H MacGibbon, J. Phys. A: Math. Gen. {\bf 17} 2581 (1984) (and references cited therein).
\bibitem{mm1} Manu Mathur and Diptiman Sen, J. Math. Phys. {\bf 42}, 4181 (2001). 
\bibitem{mu1} S. Chaturvedi and N. Mukunda, J. Math. Phys. {\bf 43}, 5262 (2002).
\bibitem{rmi} Ramesh Anishetty, Manu Mathur, Indrakshi Raychowdhury, J.\ Math.\ Phys.\ {\bf 50}, 053503 (2009). 
\bibitem{coleman} Sidney Coleman, J. Math. Phys. {\bf 5} 1343 (1964).
\bibitem{georgi} H. Georgi, {\it Lie Algebras in Particle Physics}
(Benjamin/Cummings, Reading, 1982).
\bibitem{mm2} Manu Mathur and H. S. Mani, J. Math. Phys. {\bf 43 } 5351 (2002). 
\bibitem{rmi2} R.~Anishetty, M.~Mathur and I.~Raychowdhury,
J.\ Phys.\ A {\bf 43}, 035403 (2010).
\end{thebibliography}
\end{document}